\documentclass[aps,superscriptaddress,long,notitlepage,letterpaper,balancelastpage,nofootinbib,prl,floatfix,twocolumn]{revtex4-2}
\pdfoutput=1
\usepackage{amsmath,mathtools,amssymb,amsthm,amsxtra,overpic,bbm,epsfig,url,bm}
\usepackage{hyperref}
\usepackage{mathrsfs}
\usepackage{color}
\usepackage[dvipsnames]{xcolor}
\usepackage{comment}
\usepackage{float}
\usepackage{enumitem}
\usepackage{lmodern}
\usepackage{slashed}
\usepackage{multirow}
\usepackage{appendix}
\usepackage{tabularx}
\usepackage[T1]{fontenc}  
\usepackage{caption}
\usepackage{subcaption}
\usepackage{graphicx}
\usepackage{slashed}
\usepackage{ragged2e}
\usepackage[normalem]{ulem}

\DeclareUnicodeCharacter{03BC}{\ensuremath{\mu}}
\DeclareUnicodeCharacter{2032}{\ensuremath{'}}
\definecolor{nicered}{rgb}{0.5,0.,0.}
\definecolor{nicegreen}{rgb}{0.,0.5,0.}
\definecolor{niceblue}{rgb}{0.,0.,0.5}
\hypersetup{
	colorlinks=true,
	linkcolor=black,
	filecolor=nicegreen,      
	urlcolor=niceblue,
	citecolor=nicered,
}

\newcommand{\prlsection}[2]{{\it\textbf{#1}{#2}}---}
\makeatletter
\newcommand*{\balancecolsandclearpage}{%
	\close@column@grid
	\cleardoublepage
	\twocolumngrid
}

\setlist{nolistsep}

\makeatother
\begin{document}
\preprint{N3AS-24-XXX}

\title{\vspace{1cm} \Large 
Dark population transfer mechanism for sterile neutrino dark matter
}

\author{\bf George M. Fuller}
\email[E-mail:]{gfuller@physics.ucsd.edu}
\affiliation{Department of Physics, University of California, San Diego, La Jolla, CA 92093-0319, USA}

\author{\bf Lukáš Gráf}
\email[E-mail:]{lukas.graf@:berkeley.edu}
\affiliation{Department of Physics, University of California, Berkeley, CA 94720, USA}
\affiliation{Department of Physics, University of California, San Diego, La Jolla, CA 92093-0319, USA}

\author{\bf Amol V. Patwardhan}
\email[E-mail:]{apatward@umn.edu}
\affiliation{SLAC National Accelerator Laboratory, 2575 Sand Hill Rd, Menlo Park, CA 94025, USA}
\affiliation{School of Physics and Astronomy, University of Minnesota, Minneapolis, MN 55455, USA}


\author{\bf Jacob Spisak}
\email[E-mail:]{jspisak@ucsd.edu}
\affiliation{Department of Physics, University of California, San Diego, La Jolla, CA 92093-0319, USA}
\affiliation{Max-Planck-Institut f{\"u}r Kernphysik, 69117 Heidelberg, Germany}

\begin{abstract}
We present a mechanism for producing a cosmologically-significant relic density of one or more sterile neutrinos. This scheme invokes two steps: First, a population of \lq\lq heavy\rq\rq\ sterile neutrinos is created by scattering-induced decoherence of active neutrinos; Second, this population is transferred, via sterile neutrino self-interaction-mediated scatterings and decays, to one or more lighter mass ($\sim 10\,{\rm keV}$ to $\sim 1\,{\rm GeV}$) sterile neutrinos that are far more weakly (or not at all) mixed with active species and could constitute dark matter. Dark matter produced this way can evade current electromagnetic and structure-based bounds, but may nevertheless be probed by future observations.
\noindent 
\end{abstract}
\maketitle

\prlsection{Introduction}{.}
An outstanding issue at the heart of physics and astronomy is the identity of the dark matter (DM). Among the many beyond-Standard-Model (BSM) candidates for DM, the neutrino sector offers an alluring possibility: a Standard Model (SM) singlet \lq\lq sterile\rq\rq\ neutrino. However, the simplest mechanisms for producing a dark matter-relevant relic density of these particles are challenged by X-ray and large scale structure observations (e.g., \cite{Abazajian:2001, AFT2001, Dolgov:2000ew, Kusenko:2009up, Abazajian:2009hx, Drewes:2016upu, Boyarsky:2018tvu, Dasgupta:2021ies, Abazajian:2019ejt, Zelko:2022tgf, Dolgov:2000ew} and references therein). In this paper we propose a mechanism for producing a relic density of sterile neutrinos that may be able to evade current bounds, yet may be probed by future observations. 

The standard Dodelson-Widrow mechanism \cite{Dodelson_1994} for sterile neutrino dark matter production requires only a vacuum mixing between the sterile and active (SM) neutrinos. At ultra-high temperatures in the early Universe, this mixing is medium-suppressed, but subsequently, at lower temperatures, $T\sim {\rm GeV}$, a cosmologically significant abundance of sterile neutrinos can be built-up via active neutrino scattering-induced decoherence. However, the same mixing that enables this freeze-in production scenario also provides a sterile neutrino radiative decay channel into a photon and an active neutrino~\cite{Pal:1981rm}.  Demanding a relic density of a long-lived DM candidate in this scheme picks out a $m_{\rm s} \sim {\rm keV}$ mass scale associated with the sterile state. That implies a decay-generated X-ray line at an energy $E_\gamma = m_{\rm s}/2$ that is nicely matched to the energy sensitivity range of the X-ray observatories, enabling them to provide stringent constraints \cite{AFT2001}.  

Other production mechanisms for sterile neutrino DM abound, but all require additional new physics on top of the minimalist mass and vacuum mixing parameters of Dodelson-Widrow. Some propose the sterile neutrino DM abundance to stem from decays of heavier particles~\cite{Shaposhnikov:2006xi,Petraki:2007gq,Merle:2013wta,Shuve:2014doa} or primordial black holes~\cite{Chen:2023tzd}, while being relatively agnostic about the active-sterile coupling. Similarly, the correct sterile neutrino population can be achieved via thermalization~\cite{Hansen:2017rxr}, dilution of their initially produced overabundance~\cite{Patwardhan:2015kga}, or, e.g., via SIMP-like freeze-out~\cite{Herms:2018ajr}. Other production scenarios focus on altering the active-sterile mixing. One example is the Shi-Fuller mechanism~\cite{Shi:1998km, Abazajian:2001, Kishimoto:2006zk, Kishimoto:2008ic, Ghiglieri:2015jua, Venumadhav:2015pla, Horner:2023cmc} wherein the in-medium active-sterile mixing is resonantly enhanced by a cosmic lepton number asymmetry, which may be generated through the asymmetric decay of heavier sterile states (\`a la $\nu$MSM: e.g.,~\cite{Asaka:2005an, Asaka:2005pn, Asaka:2006ek, Shaposhnikov:2006xi, Shaposhnikov:2008pf} and many more). Another invokes vacuum parameters mediated by scalar~\cite{Bezrukov:2017ike,Bezrukov:2018wvd,Bezrukov:2019mak} or axion-like~\cite{Berlin:2016bdv} fields. 

Intriguing dynamics can be achieved by bringing self-interactions into the picture. Strongly self-interacting active neutrinos can boost sterile neutrino production and, hence, result in the required DM relic abundance at smaller mixing angles~\cite{DeGouvea:2019wpf, Chichiri:2021wvw}. On the other hand, self-interactions among the sterile neutrinos themselves can lead to resonantly enhanced production, which has been investigated in the case of a  heavy mediator~\cite{Johns:2019cwc} and more recently also in the light mediator limit~\cite{Bringmann:2022aim,Astros:2023xhe}.

These considerations lead us to a scheme that utilizes the advantages of both active neutrino scattering-induced decoherence and sterile neutrino self interactions. Though there could be many sterile states, in a dark sector for example~\cite{Abazajian:2000hw}, we can illustrate the key features of our mechanism by considering sterile neutrino DM production in the presence of two sterile species with a dark self-interaction channel. This simple scenario may be further motivated by the experimental observations requiring at least two non-zero active neutrino masses. 



\prlsection{Dark Sector Dynamics}{.}
The basic idea of this mechanism is the following: (1) The heavier of the two sterile neutrinos has a small mixing with active neutrinos, enabling scattering-induced decoherence to populate this heavier sterile state; (2) Before the heavy sterile neutrino population can decay back into the SM by virtue of its mixing, self-interactions within the sterile sector engender a population transfer to the lighter sterile neutrino state. (3) The population of the lighter sterile state (or states) created this way persists until the present and can be the dark matter or a component of it, provided its mixing with the active neutrinos is sufficiently small.

The background evolution of  the SM plasma and the dark sector are governed by the following set of equations:
\begin{align}
    \frac{\mathrm{d} \rho_\text{SM}}{\mathrm{d}t} &= -3 H (\rho_{\text{SM}} + P_{\text{SM}}) - \frac{\mathrm{d} \rho_{\text{inj}}}{\mathrm{d}t} \label{eq:cosmology1}, \\
    \frac{\mathrm{d} \rho_{\text{DS}}}{\mathrm{d}t} &= -3 H (\rho_{\text{DS}} + P_{\text{DS}}) + \frac{\mathrm{d} \rho_{\text{inj}}}{\mathrm{d}t} 
    \label{eq:cosmology2}, \\
    H^2 &= \frac{8 \pi G}{3} (\rho_{\text{SM}} + \rho_{\text{DS}}),
    \label{eq:cosmology3}
\end{align}
where $\rho_\text{SM/DS}$ and $P_\text{SM/DS}$ are the energy density and pressure in the SM plasma and dark sector, respectively, and $H$ is the Hubble parameter. The first two equations describe energy conservation within each sector, with the exception of a small amount of energy transfer $\text{d} \rho_{\text{inj}}/\text{d}t$ from the SM plasma into the dark sector. The last equation is simply the Friedmann equation in a flat universe.

The energy injected into the dark sector via scattering-induced decoherence is \cite{Abazajian:2001} 
\begin{equation}
    \frac{\mathrm{d} \rho_{\text{inj}}}{\mathrm{d}t} = \int \frac{\mathrm{d}^3p}{(2 \pi)^3} E f_{\nu_\alpha}^{(\text{eq})} \frac{\Gamma_\alpha}{2} P(\nu_\alpha \rightarrow N_1), \label{eq:rhoinj}
\end{equation}
where $P(\nu_\alpha \rightarrow N_1)$ is the probability that an active neutrino of flavor $\alpha=e, \mu$, or $\tau$, energy $E$ and momentum $p$ has converted into $N_1$, the heavier sterile neutrino. It depends on the mixing between $N_1$ and $\nu_\alpha$; here we will assume that the mixing occurs only with one flavor, characterized by a single vacuum mixing angle $\theta$. While in our calculations we treat the active-sterile mixing as a free parameter, it nevertheless arise in specific models, e.g., Seesaw mass models~\cite{deGouvea:2016qpx,Cai:2017jrq}.

The scattering rate of an active neutrino on the SM plasma with a temperature $T_\text{SM}$ is 
\begin{equation} \label{eq:nuactivescatt}
    \Gamma_\alpha(p,T_{\text{SM}}) = C_\alpha(p, T_{\text{SM}}) G_F^2 p T_\text{SM}^4,
\end{equation}
where $C_\alpha(p, {T_\text{SM}})$ is a temperature-dependent coefficient and $G_F$ is the Fermi coupling constant. Assuming that the SM neutrinos follow a thermal Fermi-Dirac energy distribution with zero chemical potential for $T_{\rm SM} \gg 1 \, \text{MeV}$, one has $f_{\nu_\alpha}^{(\text{eq})}(E, T_\text{SM})= (e^{E/T_\text{SM}} + 1)^{-1}$. We ignore Pauli-blocking effects and the conversion of sterile neutrinos back into active neutrinos, thereby decoupling the energy injection rate from the dynamics within the dark sector. See the appendix for additional details. 

In general, the evolution of the sterile neutrinos is described by a set of coupled Boltzmann equations. However, if we restrict ourselves to the regime where the dark sector quickly reaches thermal equilibrium with itself as a result of self-interactions, then the dark sector's energy density and pressure are completely determined by its temperature, $T_\text{DS}$, and the masses of the dark sector constituents.  This eliminates the need to evolve the individual Boltzmann equations, and Eqs.~\eqref{eq:cosmology1}-\eqref{eq:rhoinj} form a set of coupled equations for $T_\text{DS}$, $T_\text{SM}$, and the scale factor as functions of time. To good approximation, the latter two quantities can be solved for independently of $T_\text{DS}$, since the energy injected into the dark sector is relatively small. To solve for $T_\text{DS}$, we numerically evolve Eq.~\eqref{eq:cosmology2} from a starting plasma temperature $T_\text{SM}$ at which scattering-induced decoherence is highly suppressed, with zero initial abundance for both sterile states.


\begin{figure}[t!]
    \centering
    \includegraphics[width=0.49\textwidth]{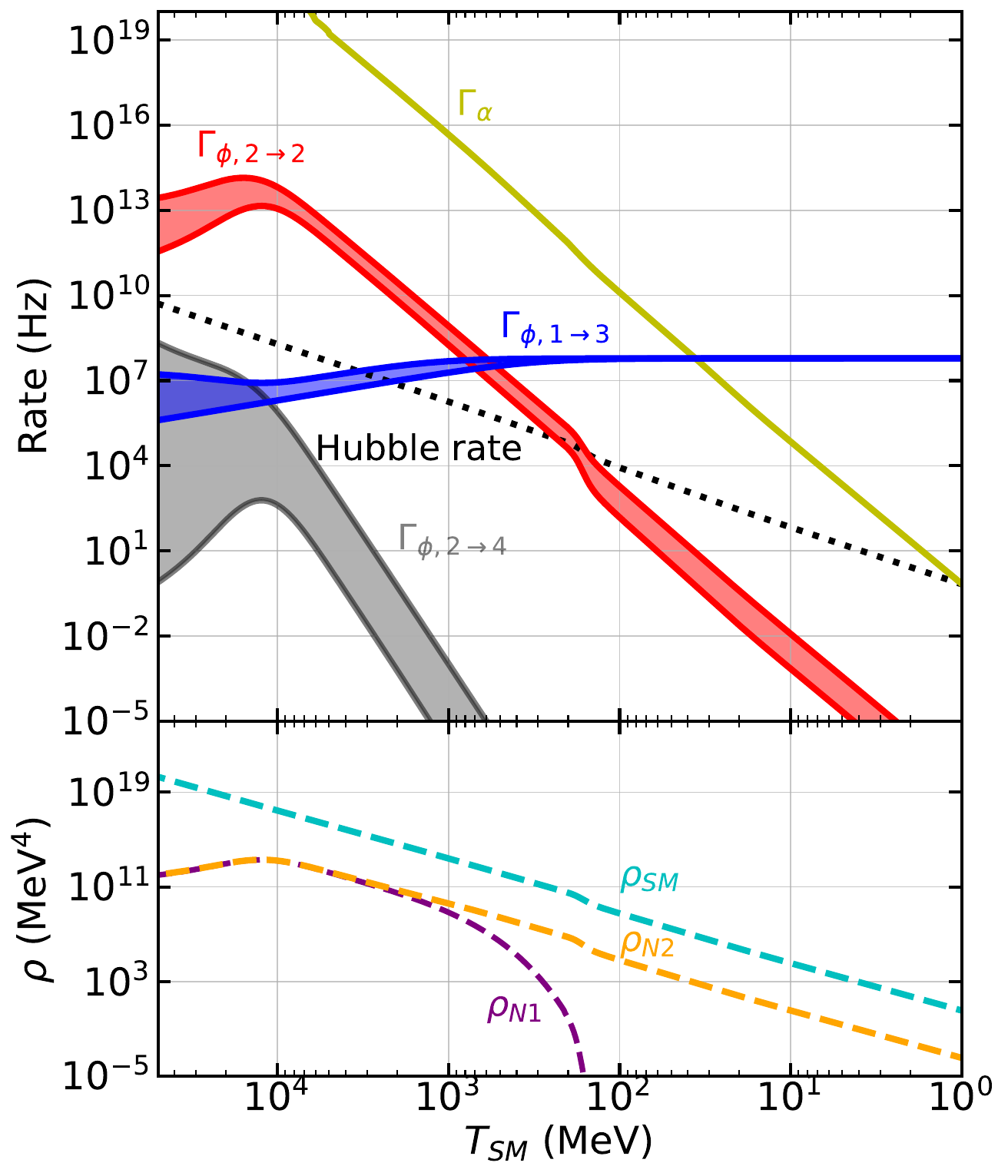}
    \caption{\justifying (\textit{Bottom panel}) Evolution of SM and sterile neutrino energy densities as a function of SM plasma temperature $T_\text{SM}$. In this scenario, thermalization within the dark sector, and population transfer from the heavier ($N_1$) to the lighter ($N_2$) sterile neutrino, are achieved through contemporaneous 2-to-2 scatterings and 1-to-3 decays.
    (\textit{Top panel}) Rates of various processes [Eqs.~\eqref{eq:cosmology3}, \eqref{eq:nuactivescatt}, and \eqref{eq:2to2}--\eqref{eq:2to4}], as indicated. The width of the colored bands represents the variation of the rates as the dark sector thermalizes (see text for details). Here, $m_{N_1} = 1\,$GeV, $\sin^2(2 \theta) = 10^{-15}$, $G_\phi = 0.1 G_F$, $\alpha=$ electron, and all rates are evaluated at $p=3T$. 
     }
    \label{fig:thermalization_1to3_example}
\end{figure}
\
To demonstrate how the mechanism could work, we choose the dark sector to consist of two sterile neutrinos $N_1$ (heavy) and $N_2$ (light) with a scalar mediator $\phi$:
\begin{equation}
\mathcal{L} \supset \frac{g^{ij}_{\phi}}{2}\overline{N}^C_j N_i \phi + \rm{h.c.},
\label{selfint}
\end{equation}
where $i,j=1,2$. For simplicity, we take all $g^{ij}_\phi=g_\phi$ to be equal, and take the heavy-mediator limit wherein $G_\phi \equiv g_\phi^2/m_\phi^2$ is the effective interaction strength. Note that taking this limit makes the whole discussion quite general, as it allows us to be agnostic about the specific UV origin of the self-interaction.

Thermal equilibrium and population transfer within the dark sector may be achieved through various combinations of self-interaction processes, namely, 2-to-2 scatterings, 1-to-3 decays, and potentially 2-to-4 scatterings. These rates scale as follows (see the appendix for a more detailed discussion): 
\begin{align}
    \Gamma_{\phi, 2 \rightarrow 2} &\propto \beta G^2_\phi p T^4 \label{eq:2to2}\\
    \Gamma_{\phi, 1 \rightarrow 3} &\propto G^2_\phi m_N^5/\gamma(p) \label{eq:1to3} \\
    \Gamma_{\phi, 2 \rightarrow 4} &\propto G^2_\phi T^4 \Gamma_{\phi, 2 \rightarrow 2}. \label{eq:2to4}
\end{align}

Here, $T$ and $\beta$ are the temperature and a suppression factor, respectively, parameterizing a Fermi-Dirac-like distribution for the sterile neutrino(s): $f(E,T) \simeq \beta / (e^{E/T} + 1)$. $\Gamma_{\phi, X \rightarrow Y}$ refers to a $\phi$-mediated process with $X$ sterile neutrinos in the initial state and $Y$ sterile neutrinos in the final state. $\gamma(p)$ in Eq.~\eqref{eq:1to3} is the Lorentz factor for a heavy sterile neutrino with momentum $p$, accounting for a time-dilation of its lifetime. 

These may be evaluated in two regimes of interest: \textit{without} and \textit{with} dark sector internal thermalization. Without thermalization, a suppressed Fermi-Dirac form with $T = T_\text{SM}$ and $\beta \ll 1$ is a reasonable approximation to the $N_1$ distribution generated by active neutrino scattering-induced decoherence. Subsequently, if the dark sector internally thermalizes, the distribution functions instead assume \textit{un-supressed} Fermi-Dirac forms (i.e., with $\beta = 1$) parameterized by a \textit{dark sector temperature} $T_\text{DS}$. Assuming an empty dark sector to begin with, and subsequent energy conservation within the dark sector as it thermalizes, one must have 
\begin{equation}\label{eq:thermalization_temp}
    g_\text{DS}^\star T_\text{DS}^4 = g_{N_1}^\star \beta T_\text{SM}^4,
\end{equation}
 and therefore $T_\text{DS} < T_\text{SM}$. Here, $g^\star_{N_1}$ and $g^\star_\text{DS}$ are respectively the number of degrees of freedom of the heavy sterile $N_1$, and in the dark sector overall. For a typical momentum $p \simeq 3T$, $\Gamma_{\phi, 2 \rightarrow 2}$ and $\Gamma_{\phi, 2 \rightarrow 4}$ are proportional to $T^5$ and $T^9$, respectively. Therefore, the decrease in temperature from $T_\text{SM}$ to $T_\text{DS}$ outweighs the increase of $\beta$ from $\ll 1$ to $\beta = 1$, leading to an overall reduction in the rates post-thermalization. The decay rate for the heavy sterile neutrino in the plasma rest frame is always larger after dark sector thermalization because $T_\text{DS} < T_\text{SM}$.

To assess the prospects of dark sector internal thermalization, these rates may be compared to the Hubble rate, $H = \sqrt{8 \pi g^\star_\text{SM} / 3} (T^2 / m_\text{pl})$, where $g^\star_\text{SM}$ is number of relativistic degrees of freedom in the SM plasma. On the other hand, to avoid resonant overproduction due to the additional in-medium potential provided by the self-interaction, we require that $\Gamma_{\phi, 2 \rightarrow 2} < \Gamma_\alpha$ \cite{Johns:2019cwc}. We also require that $g_{\phi} < 1$ to stay in the perturbative regime. These requirements may be simultaneously satisfied over some range of $G_\phi$ values, given a sterile neutrino mass $m_{N_1}$ and vacuum mixing angle $\theta$.

This mechanism is illustrated for a particular choice of parameters in Fig.~\ref{fig:thermalization_1to3_example}. For both $\Gamma_{\phi, 2 \rightarrow 2}$ and $\Gamma_{\phi, 2 \rightarrow 4}$, the upper and lower edges of each band reflect the rate expressions without and with dark sector internal thermalization, respectively. For $\Gamma_{\phi, 1 \rightarrow 3}$ the situation is reversed. $N_1$ is produced via scattering-induced decoherence, and initially the dark sector has not achieved internal thermal equilibrium due to a lack of number-changing processes. This changes once $\Gamma_{\phi, 1 \rightarrow 3} > H$: this process increases the number of sterile neutrinos, and the 2-to-2 processes, which are initially much faster than the decays, ensure the $N_1$ and $N_2$ populations equilibrate with roughly equal number densities. Note that SM decay branches for $N_1$ are subdominant for these parameter choices compared to the dark decays (see the appendix). As the dark sector thermalizes, the rates switch to the opposite sides of the bands\footnote{Unless a freeze-out ($\Gamma_\phi < H$) occurs in the midst of thermalization, in which case, the rate for that process never quite reaches the opposite edge of the band.}. Eventually, $N_1$ becomes non-relativistic and is Boltzmann-suppressed, transferring $N_1$ into $N_2$. 1-to-3 decays, which eventually become faster than 2-to-2 processes, further aid this transfer. $N_2$ then persists until the present day as a component of dark matter.

An alternative scenario can achieve this dark sector internal thermalization through 2-to-4 processes. Although these are relevant only at high temperatures and for a much smaller slice of the parameter space than the previous example, they would allow for smaller $N_1$ masses. This is because $\Gamma_{\phi, 1 \rightarrow 3} \propto N_1^5$, and so smaller masses have a lower 1-to-3 rate that may not be significant at high enough $T_{\text{SM}}$ to overlap with 2-to-2 processes.

\begin{figure*}[htbp]
  \centering
  \subfloat[]{\includegraphics[width=0.48\textwidth]{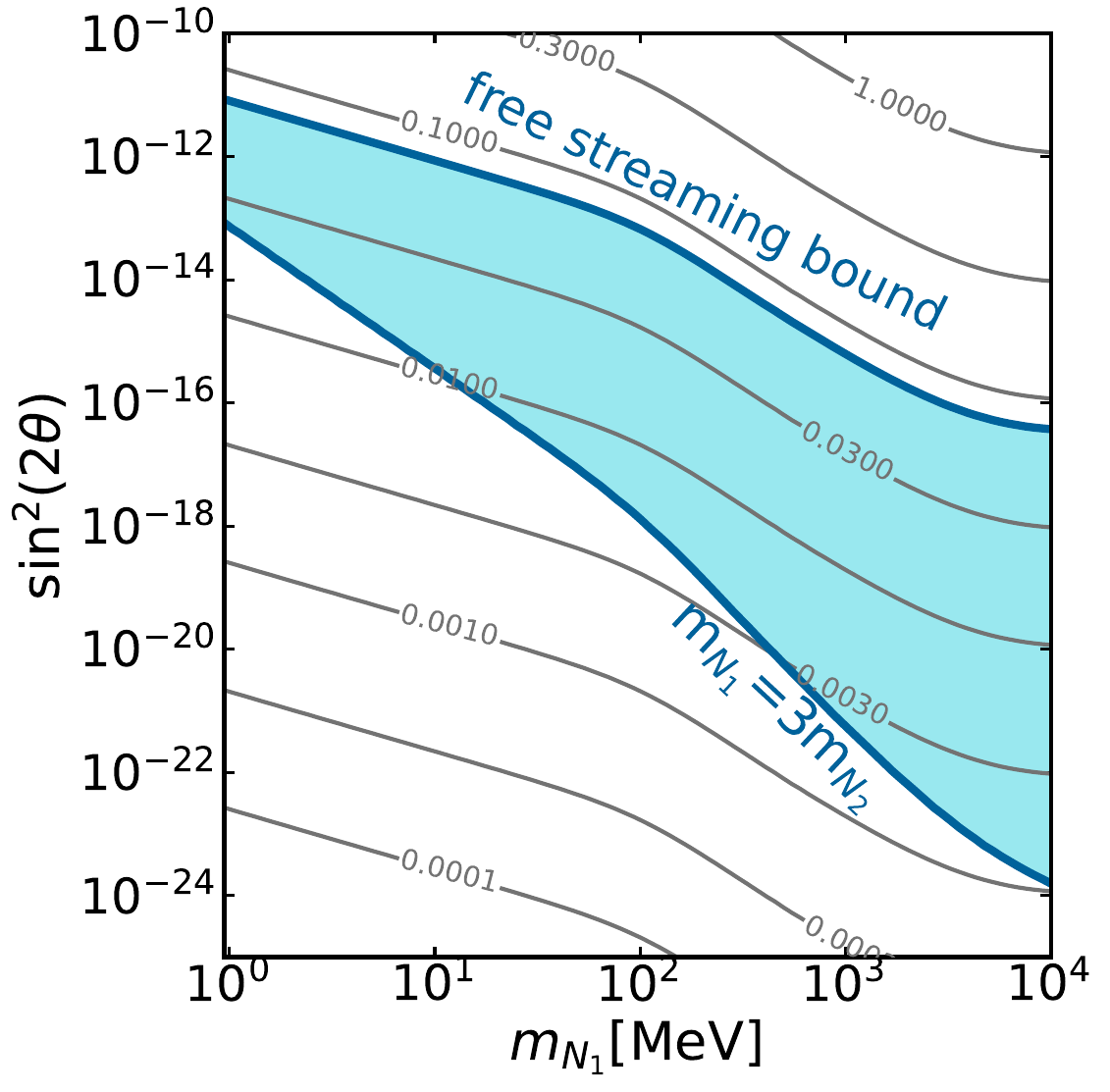}\label{fig:f1}}
  \hfill
  \subfloat[]{\includegraphics[width=0.48\textwidth]{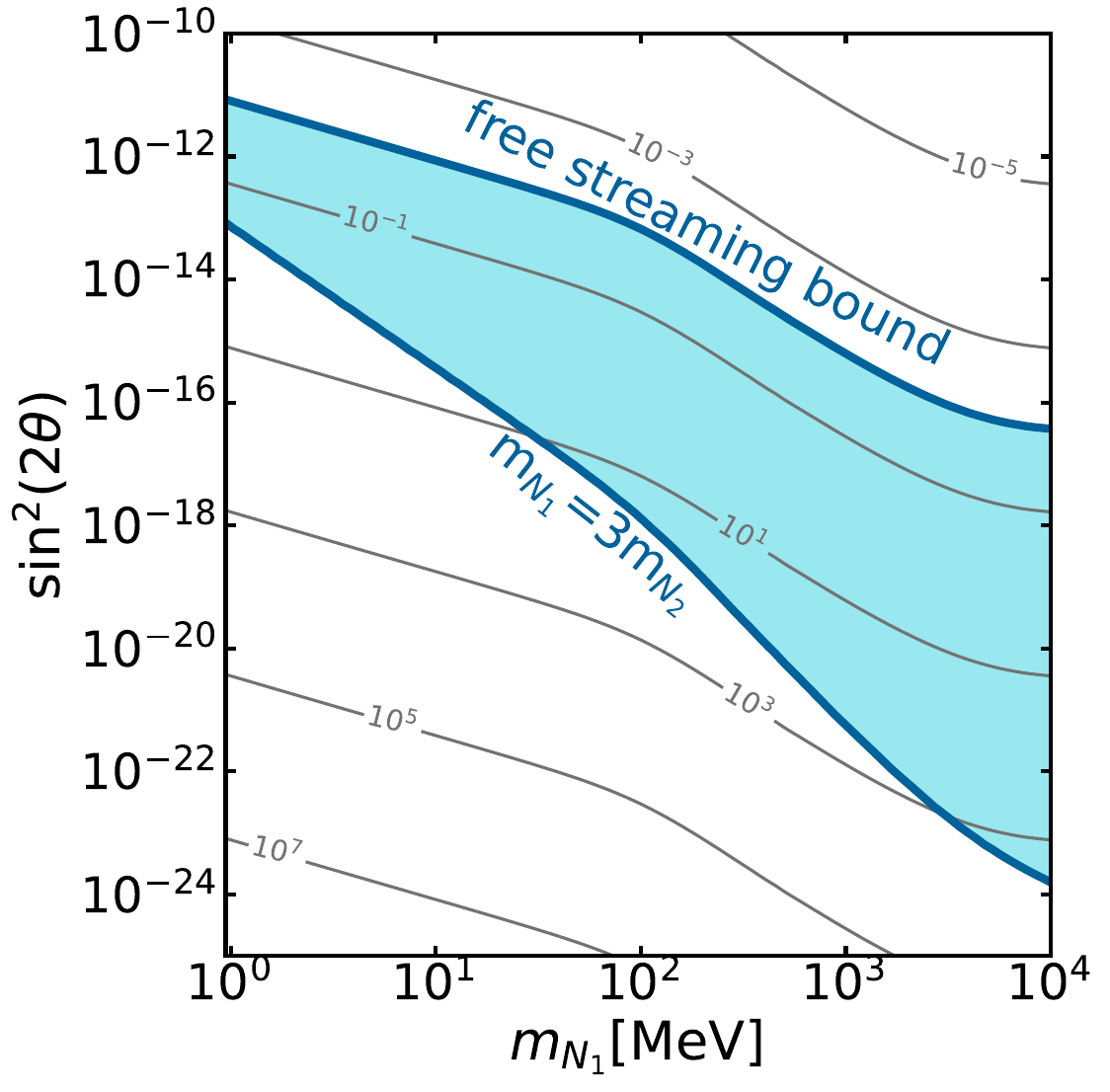}\label{fig:f2}}
  \hfill
  \caption{\justifying
  Across a parameter space of the heavy sterile neutrino rest-mass ($m_{N_1}$) and its mixing angle with the active neutrinos ($\sin^2(2\theta)$), the following contours are shown. \textit{Left}: The ratio of the dark sector temperature to the standard model temperature, i.e., $T_\text{DS}/T_\text{SM}$, stemming from the energy injected into the dark sector from scattering-induced decoherence, under the assumption that the dark sector thermalizes with itself. The blue region is where all observational and theoretical bounds are satisfied.
  \textit{Right}: The mass of $N_2$ (in MeV) required to account for all of the dark matter in this scenario.
  }
  \label{fig:results}
\end{figure*}

\prlsection{Parameter Space}{.}
Under the assumption that the dark sector thermalizes internally, the parameter space in which all of the dark matter is produced is shown in Fig.~\ref{fig:results}. The left panel shows contours of the final dark sector temperature relative to the SM temperature. In the viable parameter space this ratio is less than one for two reasons: the active-sterile mixing is too small to thermalize the dark sector with the SM, and the SM plasma is heated by the disappearance of relativistic degrees of freedom as the temperature drops. The right panel of Fig.~\ref{fig:results} shows contours of the required mass $m_{N_2}$ necessary to make all of the dark matter, given the dark sector temperatures shown in the left panel. It is assumed that the $N_1$ relic population is converted to $N_2$ via {2-to-2 scattering}, although the results do not qualitatively change as long as the number density is roughly preserved, e.g., if 1-to-3 decays occur instead.

The lower boundary on the colored region in the parameter space shown in Fig.~\ref{fig:results} comes from the condition that $N_1$ have a large enough mass to decay into three $N_2$, i.e. $m_{N_1} > 3 m_{N_2}$, which is the regime we restrict ourselves to in this work. The upper boundary of the color region follows from dark matter free streaming considerations, explained below. We choose the upper limit of the $N_1$ mass range in Fig.~\ref{fig:results} such that $T_\text{SM} \gg m_{N_1}$ during the scattering-induced decoherence production period, ensuring $N_1$ is produced relativistically.

\prlsection{Discussion}{.}
A key feature of this mechanism is that it allows for a larger sterile neutrino dark matter mass. Normally, the mass of the sterile neutrino is bounded from above by the requirement that the dark matter does not decay too fast via its active-sterile mixing. In the proposed mechanism, however, the mixing angle of the dark matter candidate sterile neutrino does not affect its relic abundance, and can therefore be arbitrarily small. On the flip side, this mechanism provides discovery potential for current and future X-ray and $\gamma$-ray missions across a much wider range of sterile neutrino masses and mixings. Depending on the mass of $N_1$, the dark matter mass $m_{N_2}$ in our mechanism can range from $\sim 10\,\text{keV} \text{--} 1\,\text{GeV}$. Many X-ray telescopes have explored some of this parameter space (e.g.,~\cite{Boyarsky:2005us,Riemer-Sorensen:2006uyy,Loewenstein:2008yi,Watson:2011dw,Borriello:2011,Horiuchi:2013noa,Tamura:2014mta,Hofmann:2016urz,Hofmann:2019ihc,Foster:2021ngm}), which will be further probed by the upcoming telescopes ATHENA~\cite{Nandra:2013} and XRISM~\cite{xrism:2020}. This also motivates searches for dark matter lines at higher energies, beyond the typical range for sterile neutrino dark matter, using telescopes such as NuSTAR~\cite{Harrison:2013,Riemer-Sorensen:2015kqa, Perez:2016tcq, Ng:2019gch, Roach:2019ctw, Roach:2022lgo}, HEX-P~\cite{Madsen:2023}, Fermi-GBM~\cite{Ng:2015gfa} and INTEGRAL-SPI~\cite{Yuksel:2007xh,Boyarsky:2007ge}. Moreover, heavier $N_2$ masses allow for decay channels involving additional particles such as electrons, muons, and pions \cite{Fuller:2011}, which may generate a more complicated electromagnetic signature than a narrow line.

On the opposite end, the lower bound on the sterile neutrino DM mass is set by the requirement that free-streaming dark matter should not wash out small-scale structure, that is, its average kinetic energy must not be too large. {Since dark sector internal thermalization ensures $T_\text{DS} < T_\text{SM}$, the sterile neutrino average energy can be considerably lower than in a purely scattering-induced decoherence scenario, resulting in a weaker lower mass bound}. For a thermally equilibrated species accounting for all of the dark matter, since $\rho_\text{DM} \sim g^\star m T_\text{DS}^3$, fixing the mass $m$ and the number of degrees of freedom $g^\star$ also fixes the temperature and therefore the energy spectrum. Small-scale structure observations can therefore place a lower limit on the mass of a thermal dark matter candidate. The current bound is $9.7\,\mathrm{keV}$ at $95\%$ C.L. ~\cite{Nadler_2021}. For comparison, the corresponding bound for the Dodelson-Widrow scenario can be as strong as $92 \,\mathrm{keV}$ \cite{Zelko:2022tgf}.



The heavy sterile neutrino may in principle be detected in laboratory searches, but unfortunately the mixing angles shown in Fig.~\ref{fig:results} are too small for any current or upcoming experiments to detect (see~Ref.~\cite{Bolton:2019pcu} for a review). Furthermore, a number of such experiments rely on detecting the sterile neutrino's SM decay products, which would not be effective in the scenario considered here. However, a possibility that may deserve future exploration is the case where decays into SM products dominate over the dark decay branch. This would necessitate larger mixing angles in order to produce the observed relic abundance, potentially allowing for detection in experiments close to the preferred parameter space shown in Fig.~\ref{fig:results}, such as HUNTER~\cite{Martoff:2021vxp} and TRISTAN~\cite{KATRIN:2018oow,KATRIN:2022spi}. Such scenarios could have significant implications for neutrino decoupling and big-bang nucleosynthesis (BBN)~\cite{Dolgov:2000jw, Fuller:2011, Gelmini:2020ekg, Rasmussen:2021kbf, Abazajian:2023reo}, pre-BBN cosmology~\cite{Gelmini:2019clw,Gelmini:2019esj,Gelmini:2019wfp,Chichiri:2021wvw}, as well as energy transport in core-collapse supernovae~\cite{Fuller:2008erj,Chauhan:2023sci}. The lighter sterile neutrino $N_2$, if it also mixes with the active neutrinos, could be resonantly produced in a supernova core, affecting the explosion dynamics~\cite{Hidaka:2006sg,Hidaka:2007se,Arguelles:2016uwb}. To faithfully assess the impact of sterile neutrinos in these environments, a careful consideration of dynamical feedback effects in their production is important~\cite{Suliga:2020vpz,Ray:2023gtu}.

A possible direction for future work could be to explore the case of a lighter dark sector mediator. This would result in more complicated dynamics, including resonances in scattering rates and the possibility of producing the mediator itself. In fact, a lighter mediator with just one sterile neutrino can produce the correct relic abundance at some masses by modifying the typical scattering-induced decoherence production process~\cite{Bringmann:2022aim,Astros:2023xhe}. Sterile neutrinos with strong self-interactions mediated by lighter scalar fields may reproduce the regimes preferred by scenarios of self-interacting dark matter~\cite{Spergel:1999mh, Rocha:2012jg, Tulin:2017ara, Garcia-Cely:2017qpx, Mavromatos:2017lfs, Adhikari:2022sbh} addressing the small structure problems~\cite{Moore:1994yx,Oh:2008ww,Oh:2015xoa}. 
Moreover, for certain masses and mixings, a resonance in the sterile neutrino scattering rate could cause dips in the diffuse supernova neutrino background~\cite{Suliga:2023}.

The phenomenology of sterile neutrino dark matter with additional particle species and/or secret interactions in the dark sector can be quite rich. An extension of this mechanism may include multiple light sterile species. This could correspond, for example, to a tower of massive Kaluza-Klein modes~\cite{Abazajian:2000hw}, or other rich dark sectors. If the heaviest of these species is the only one produced in significant abundance through scattering-induced decoherence, its decays into the lighter sterile neutrinos may lead to an arbitrarily cold dark matter spectrum (i.e. for arbitrarily large $g_{\text{DS}}^\star$, see Eq.~\eqref{eq:thermalization_temp}). Such a scenario also would potentially provide a multitude of targets, at different masses and mixing angles, for future electromagnetic searches. We conclude that the future of this dark subject could be bright.


\prlsection{Acknowledgments}{.}
The authors would like to thank Kev Abazajian, Chad Kishimoto, Richard Rothschild and Anna Suliga for fruitful discussions. This work was supported in part by National Science Foundation (NSF) grant PHY-2209578 at UCSD, the {\it Network for Neutrinos, Nuclear Astrophysics}, and Symmetries (N3AS), funded by NSF grants  PHY-2020275 and PHY-1630782, and the Heising-Simons Foundation, grant 2017-228. AVP would like to acknowledge support from the U.S. Department of Energy (DOE) under contract number DE-AC02-76SF00515 at SLAC National Accelerator Laboratory and DOE grant DE-FG02-87ER40328 at the University of Minnesota. J. Spisak 
 acknowledges support from the UCSD Academic Senate.  We would also like to acknowledge the Kavli Institute for Theoretical Physics (NSF grants PHY-1748958 and PHY-2309135), the Institute for Nuclear Theory (DOE grant DE-FG02-00ER41132), and the Mainz Institute for Theoretical Physics (MITP), Cluster of Excellence PRISMA+ (Project ID 39083149). The hospitality of these institutions and programs was vital for this work.

\appendix

\section{Appendix}

\prlsection{Scattering-Induced Decoherence}{.}
The purpose of this section is to explain how the energy injection rate from the SM plasma into the dark sector is calculated.

The Boltzmann equation that describes the production of a sterile neutrino $N$ via scattering-induced decoherence due to mixing with a single active species $\alpha$, in the absence of any sterile neutrino self-interaction, is \cite{Abazajian:2001}
\begin{equation}\label{eq:boltzmann}
    \left( \frac{\partial}{\partial t} - H p \frac{\partial}{\partial p} \right) f_N = \frac{1}{2} \Gamma_\alpha P(\nu_\alpha \rightarrow N) (f_{\nu_\alpha} - f_N).
\end{equation}

$f_N$ and $f_{\nu_\alpha}$ are the sterile and active neutrino distribution functions, respectively, and the other variables are as described in the main text. All variables on the RHS depend on time and momentum. The probability to transform an active neutrino into a sterile neutrino is 

\begin{widetext}
\begin{equation}
    P(\nu_\alpha \rightarrow N) =
    \frac{1}{2} \frac{\Delta^2 \sin^2(2 \theta)}{\Delta^2 \sin^2(2 \theta) + \left(\frac{\Gamma_\alpha}{2}\right)^2 + (\Delta \cos(2 \theta) - V_\alpha)^2}.
\end{equation}
\end{widetext}

Here $\Delta \approx \frac{m_N^2}{2p}$, $\theta$ is the vacuum active-sterile mixing angle, and $\left( \frac{\Gamma_\alpha}{2} \right)^2$ is the quantum damping factor. The key inputs to the calculation, besides the sterile neutrino mass and mixing angle, are the active neutrino scattering rate $\Gamma_\alpha$ and in-medium potential $V_\alpha$, both of which depend on the neutrino momentum $p$ and SM plasma temperature $T_{\text{SM}}$.

The active neutrino scattering rate for $T_\text{SM} \in$ [1 MeV, 10 GeV] is taken from \href{http://www.laine.itp.unibe.ch/neutrino-rate/}{publicly available data files} described in section 3 of reference \cite{Asaka:2006nq}. For $T_\text{SM} \in$ [5 GeV, 150 GeV], rates are taken from \href{http://www.laine.itp.unibe.ch/production-midT/}{publicly available data files} described in section 2 of reference \cite{Ghiglieri_2016}. Note that for this temperature range the approximate scaling $\Gamma_\alpha \propto p T^4_\text{SM}$ breaks down and $C_\alpha(p, T_{\text{SM}})$ varies by several orders of magnitude.

The in-medium potential (with negligble lepton number) for 1 MeV $\ll T_{\text{SM}} \ll T_{\text{EW}}$, where $T_{\text{EW}}$ is the electroweak scale, is \cite{Notzold:1987ik}:

\begin{equation}\label{eq:matter_potential_below_EW}
\begin{split}
    V_{\alpha}(p,T) = &- \frac{8 \sqrt{2} G_F p}{3 m_Z^2} ( n_{\nu_\alpha} \langle E_{\nu_\alpha} \rangle + n_{\overline{\nu}_\alpha} \langle E_{\overline{\nu}_\alpha} \rangle) \\
    &- \frac{8 \sqrt{2} G_F p}{3 m_W^2} ( n_{l_\alpha} \langle E_{l_\alpha} \rangle + n_{\overline{l}_\alpha} \langle E_{\overline{l}_\alpha} \rangle).
\end{split}
\end{equation}

Eq.~\eqref{eq:matter_potential_below_EW} represents a leading-order expansion of the W and Z propagators, which works well for calculations involving keV sterile neutrino masses because the production rate peaks around a SM temperature $T_\text{peak} \approx 133~\text{MeV} \left( \frac{m_N}{1~\text{keV}} \right)^\frac{1}{3}$ \cite{Abazajian:2001}. For larger masses, production occurs at higher temperatures and the leading-order expansion is no longer a good approximation. In fact, the potential is positive above the EW scale, and the sign change induces a resonance which enhances the production rate \cite{Alonso_Alvarez_2022}. Above the EW scale, the potential takes the form \cite{Quimbay_1995}

\begin{equation} \label{eq:matter_potential_above_EW}
    V_a = (3 g^2 + g'^2) T_{\text{SM}}^2/(32 p),
\end{equation}

where $g$ and $g'$ are the SU(2)$_L$ and U(1)$_Y$ SM gauge couplings, respectively. The two regimes in equations \ref{eq:matter_potential_below_EW} and \ref{eq:matter_potential_above_EW} may be smoothly connected by the full integral form of the potential described in appendix A of reference \cite{Alonso_Alvarez_2022}, which we summarize below. The full form of the potential after EW symmetry breaking is

\begin{equation}\label{eq:full_below_EW}
\begin{split}
    V_{\alpha}(p,T) = - \pi \alpha_w (&B(0, m_Z)/\cos^2(\theta_W) + \\&B(m_{l_\alpha}, m_W) (2+m_{l_\alpha}^2/m_W^2)).
\end{split}
\end{equation}

Here $m_{l_\alpha}$ is the mass of the charged lepton of flavor $\alpha$, $\theta_W$ is the electroweak mixing angle, and 
\begin{equation}
\begin{split}    
    B(m_f, m_A) &= \frac{1}{p^2} \int \frac{dk}{8 \pi^2} \biggl[ \left( \frac{k \delta m^2 }{2 E_A}L^+_2(k) - \frac{4 p k^2}{E_A} \right) f_{B}(E_A) \\
    &+ \left( \frac{k \delta m^2 }{2 E_f}L^+_1(k) - \frac{4 p k^2}{E_f} \right) f_{F}(E_f) \biggl].
\end{split}
\end{equation}
Here $\delta m^2 = m_A^2 - m_f^2$, $f_{B/F}(E_i)$ are Bose-Einstein or Fermi-Dirac distribution functions with energy $E_i = \sqrt{k^2 + m_i^2}$, and
\begin{align}
    L_{1/2}^+(k) = \ln \left( \left\lvert \frac{[\delta m^2 + 2 p (k + E_{f/A})][\delta m^2 + 2 p (k - E_{f/A})]}{[\delta m^2 - 2 p (k + E_{f/A})][\delta m^2 - 2 p (k - E_{f/A})]} \right\rvert \right)
\end{align}
The transition from Eq.~\eqref{eq:full_below_EW} to Eq.~\eqref{eq:matter_potential_above_EW} may be achieved by taking $m^2_{Z,W}(T) = m^2_{Z,W}(1-T^2/T^2_{\text{EW}})$. At temperatures far below $T_{\text{EW}}$, Eq.~\eqref{eq:full_below_EW} reduces to Eq.~\eqref{eq:matter_potential_below_EW}.

With the addition of a sterile neutrino self-interaction, the Boltzmann equation for the heavy sterile neutrino may be written as \cite{Johns:2019cwc}
\begin{equation}\label{eq:sterile_boltzmann}
    \left( \frac{\partial}{\partial t} - H p \frac{\partial}{\partial p} \right) f_{N_1} = \frac{1}{2} \Gamma_{\text{tot}} P(\nu_\alpha \rightarrow N_1) (f_{\nu_\alpha} - f_{N_1}) + \mathcal{C}_s.
\end{equation}
The total scattering rate is $\Gamma_{\text{tot}} = \Gamma_\alpha + \Gamma_{\phi, 2 \rightarrow 2}$, a collision term $\mathcal{C}_s$ describing sterile-only processes has been added, and the potential also gains an additional term from the self-interaction. Furthermore, there is an additional Boltzmann equation for $N_2$ coupled to Eq.~\eqref{eq:sterile_boltzmann}. However, two important facts allow us to ignore these complications: 1) we are in the regime where $\Gamma_{\phi, 2 \rightarrow 2} \ll \Gamma_\alpha$ (in order to avoid resonant overproduction \cite{Johns:2019cwc}), and 2) the dark sector never comes close to being in thermal equilibrium with the SM plasma. The former means that the sterile neutrino contributions to the total scattering rate and potential can be ignored. The latter means we may ignore the impact of the dark sector on the SM and conversions from sterile neutrinos back into active neutrinos, ie $f_{\nu_\alpha} - f_N$ may be replaced by $f_{\nu_\alpha}^{(\text{eq})}$. Therefore the energy injection rate from the SM into the dark sector takes the form shown in the main text, and is decoupled from the dynamics of the dark sector, which simplifies the analysis of the dark sector considerably.

\prlsection{Sterile Neutrino Interaction Rates}{.}
The purpose of this section is to explain the values of the sterile neutrino self-interaction rates used in the main text. 

The relativistic 2-to-2 sterile neutrino scattering rate is taken to be \cite{Johns:2019cwc}
\begin{equation}
    \Gamma_{\phi, 2 \rightarrow 2} \approx 0.03 \beta G^2_\phi p T^4.
\end{equation}
This rate assumes that the sterile neutrino distribution functions are Fermi-Dirac-like $f(E,T) \simeq \beta / (e^{E/T} + 1)$, ignores Pauli blocking, and uses average thermal momentum values $p \approx 3.15 T$. 

The 1-to-3 decay rate ($N_1 \rightarrow 3 N_2)$ is 
\begin{equation}
    \Gamma_{\phi, 1 \rightarrow 3} = \frac{G^2_\phi m_{N1}^5}{1024 \pi^3}
\end{equation}
in the rest frame of $N_1$ and the limit where $m_{N2}$ is set to zero. To determine if the decay channels enabled by active-sterile mixing are important, this may be compared to the rate for sterile neutrino decay into three \textit{active} neutrinos via active-sterile mixing \cite{Barger_1995},  
\begin{equation}
    \Gamma_{3 \nu_{\text{active}}} = \frac{G^2_F m_{N1}^5}{192 \pi^3} \sin^2(\theta). 
\end{equation}
For larger $m_{N_1}$ other decay channels enabled by active-sterile mixing come into play \cite{Fuller:2011}, but for $m_{N1} \leq 10 \, \text{GeV}$ the total decay rate is always $\lesssim 20 \Gamma_{3 \nu_{\text{active}}}$ \cite{Deppisch_2018}. Decays into SM products, therefore, are subdominant to the dark sector decay $\Gamma_{\phi, 1 \rightarrow 3}$ if $\sin^2(\theta) G_F^2/G_\phi^2 \ll 10^{-2}$. Since the preferred mixing angles we consider are $\sin^2(2 \theta) < 10^{-10}$, this is satisfied as long as $G_\phi$ isn't several orders of magnitude smaller than $G_F$.  

We \textit{approximate} the 2-to-4 rate via naive dimensional analysis as 
\begin{equation}
    \Gamma_{\phi, 2 \rightarrow 4} \sim G^2_\phi T^4 \Gamma_{\phi, 2 \rightarrow 2}.
\end{equation}
It is important to note that an accurate calculation of the numerical prefactor would be required in order to definitively establish the parameter space in which full thermalization is achieved through 2-to-4 processes (rates for some scalar-mediated interactions involving >4 neutrinos, including 4-to-2 annihilation, were calculated in reference \cite{Herms:2018ajr}). If this scenario is of particular interest, a more detailed rate calculation may be a fruitful topic for future work.

\bibliography{refs}
\end{document}